\documentclass[10pt,a4paper,english,prd,nofootinbib,notitlepage,superscriptaddress,showkeys,preprintnumbers,longbibliography]{revtex4-1}
\pdfoutput=1
\usepackage[T1]{fontenc}
\usepackage[latin9]{inputenc}
\usepackage{color}
\usepackage{babel}
\usepackage{array}
\usepackage{multirow}
\usepackage{amsmath}
\usepackage{amssymb}
\usepackage{graphicx}
\usepackage{esint}
\usepackage[unicode=true,
 bookmarks=false,
 breaklinks=false,pdfborder={0 0 1},backref=false,colorlinks=true]
 {hyperref}
\hypersetup{
 citecolor=blue,urlcolor=blue,linkcolor=blue}
\usepackage{breakurl}

\makeatletter

\providecommand{\tabularnewline}{\\}

\usepackage{babel}
\usepackage{babel}
\usepackage{babel}
\usepackage{babel}
\usepackage{babel}

\usepackage{color}
\usepackage{babel}
\usepackage{multirow}

\pdfpageheight\paperheight
\pdfpagewidth\paperwidth

\providecommand{\tabularnewline}{\\}

\@ifundefined{textcolor}{}{
 \definecolor{BLACK}{gray}{0}
 \definecolor{WHITE}{gray}{1}
 \definecolor{RED}{rgb}{1,0,0}
 \definecolor{GREEN}{rgb}{0,1,0}
 \definecolor{BLUE}{rgb}{0,0,1}
 \definecolor{CYAN}{cmyk}{1,0,0,0}
 \definecolor{MAGENTA}{cmyk}{0,1,0,0}
 \definecolor{YELLOW}{cmyk}{0,0,1,0}
}

\usepackage{babel}

\usepackage{breakurl}
\usepackage{cleveref}
\usepackage{ulem}
\usepackage{bbm}

\usepackage{todonotes}
\definecolor{lightblue}{HTML}{A9D0F5}
\definecolor{lightgreen}{HTML}{BCF5A9}
\definecolor{lightred}{HTML}{F6CECE}
\definecolor{lightorange}{HTML}{FFA800}
\definecolor{greengray}{HTML}{5C9393}
\definecolor{lightgreengray}{HTML}{80CCCC}

\include{aas_macros.sty}

\providecommand{\tabularnewline}{\\}

\makeatother

\begin{document}

\title{Structure formation in the Deser-Woodard nonlocal gravity model:
a reappraisal}

\author{Henrik Nersisyan}
\email{h.nersisyan@thphys.uni-heidelberg.de}

\selectlanguage{english}%

\author{Adrian Fernandez Cid}
\email{fernandez@thphys.uni-heidelberg.de}

\selectlanguage{english}%

\author{Luca Amendola}
\email{l.amendola@thphys.uni-heidelberg.de}

\selectlanguage{english}%

\affiliation{Institut f{\"u}r Theoretische Physik, Ruprecht-Karls-Universit{\"a}t
Heidelberg, Philosophenweg 16, 69120 Heidelberg, Germany}

\begin{abstract}
In this work, we extend previous analyses of the structure formation
in the $f(\Box^{-1}R)$ model of nonlocal gravity proposed by Deser
and Woodard (DW), which reproduces the background expansion of $\Lambda$CDM
with no need of a cosmological constant nor of any dimensional constant
beside Newton's one. A previous analysis based on redshift-space distortions
(RSD) data concluded that the model was ruled out. In this work we
revisit the issue and find that, when recast in a localized model,
the DW model is not ruled out and actually gives a better fit to RSD
data than $\Lambda$CDM. At the same time, the model predicts a slightly
lower value of $\sigma_{8}$ than $\Lambda$CDM, in agreement with
recent estimates based on lensing. We also produce analytical approximations
of the two modified gravity functions and of $f\sigma_{8}(z)$ as
a function of redshift. Finally, we also show how much the fit depends
on initial conditions when these are generalized with respect to a
standard matter-dominated era. 
\end{abstract}

\keywords{modified gravity, nonlocal gravity, dark energy, background cosmology}

\date{\today}
\maketitle

\section{Introduction}

The late-time accelerated expansion of the universe \cite{Riess:1998cb,Perlmutter:1998np,Sherwin:2011gv,Dunkley:2008ie,Komatsu:2008hk,vanEngelen:2012va,Scranton:2003in,Sanchez:2012sg}
is attributed in the standard cosmological model, or $\Lambda$CDM,
to the influence of dark energy in the form of a cosmological constant
$\Lambda,$ interpreted as the energy density of the vacuum. However,
this otherwise formally and observationally consistent model carries
two unsolved puzzles: the so-called coincidence and the fine-tuning
problems. The former issue refers to $\Lambda$CDM not explaining
the fact that the accelerated phase in the expansion began only recently
in the cosmological time, while the latter expresses the enormous
disagreement between the energy scale introduced by $\Lambda$ and
the predictions of the Standard Model of particle physics for the
vacuum energy density. Consequently, a wealth of alternative, more
complicated cosmological models are continuously developed and proposed
with the purpose of providing a more accurate and robust description
of our universe, the majority of which may be classified as dark energy
(if they introduce new matter content) or modified gravity (if they
depart from Einstein's general relativity) models (although of course
from a purely gravitational point of view there is no fundamental
distinction between these two classes). Typically, these new models
are required to emulate the background expansion history of the universe
given by $\Lambda$CDM, well supported by the data. The imposition
of this condition is called the reconstruction problem. Once this
step is fulfilled, one can observationally distinguish among models
by looking at their predictions beyond the background, such as solar
system tests and the structure formation in the universe. 

Within the class of modified gravity models, nonlocal gravity theories
have recently gained remarkable interest. In this direction, pioneering
works are \cite{deser_nonlocal_2007,deffayet_reconstructing_2009},
where the authors attempt to construct a viable alternative to the
standard $\Lambda$CDM cosmology through nonlocal modifications of
the form $f(\Box^{-1}R)$. This model, at the background level, has
the advantage, over $\Lambda$CDM, that it exactly reproduces the
same evolution without introducing a new energy scale. The price to
pay is the loss of structural simplicity. Indeed, in order to exactly
duplicate the $\Lambda$CDM behavior, the function $f(\Box^{-1}R)$
must be of a somewhat contrived form  \cite{deffayet_reconstructing_2009}.
On a phenomenological basis, the DW nonlocal gravity model has been
shown to be ghost-free\footnote{The localized version of the DW model has been shown~\cite{Koivisto:2008dh,Nojiri:2010pw} to be ghost-free only when the function $f(\Box^{-1}R)$ satisfies a particular ghost-freeness conditions.} and close to GR in gravitationally-bound systems \cite{Deser:2013uya}.
The behavior of the model at the perturbation level was studied in
\cite{Koivisto:2008dh} and in \cite{park_structure_2013,dodelson_nonlocal_2014}.
The authors of the last two papers found that, according to the redshift-space
distortions (RSD) observations available at the time, the DW model
was disfavored over $\Lambda$CDM by $7.8\sigma$.

In this work we revisit this problem and show that the localized version
of DW model shows a different picture according to which the DW model
is not anymore disfavored over $\Lambda$CDM and actually gives a
significantly better fit to the RSD data. At the same time, the model
predicts a slightly lower value of $\sigma_{8}$ than $\Lambda$CDM,
in agreement with recent lensing results \cite{Ade:2015rim,Hildebrandt:2016iqg}.
It is important to remark that once the background is fixed to reproduce
$\Lambda$CDM, no more free parameters are left to adjust to the RSD
data. Our results disagree with those in \cite{park_structure_2013,dodelson_nonlocal_2014}.
Despite intensive testing, we have been unable to identify the reasons
for this discrepancy; we discuss some conjectures below.

We also make one step further and relax the model-dependent assumptions
implicit in previous works concerning the initial conditions for the
perturbation equation in the matter era. More precisely, we allow
the two initial conditions for the linear growth equation to vary
(as opposed to fixing them to their standard CDM values). As we will
show, however, this improves the fit only marginally.

Throughout the paper, we work in flat space and natural units, i.e.
units such that $c=\hbar=1$.

\section{The Model}

\label{sec:model}

In Ref.~\cite{deser_nonlocal_2007} the authors proposed a model
in which the Einstein-Hilbert action is nonlocally modified as 
\begin{equation}
S_{DW}=\frac{1}{16\pi G}\int d^{4}x\ \sqrt{-g}R\bigg[1+f(\square^{-1}R)\bigg],\label{action}
\end{equation}
where the nonlocal distortion function $f$ is a free function of
the inverse d'Alembertian acting on the Ricci scalar, $\Box^{-1}R$.
Since this combination is dimensionless, the Lagrangian does not introduce
any new energy scale. Variation of (\ref{action}) with respect to
the metric $g_{\mu\nu}$ yields the modified Einstein equations, 
\begin{equation}
G_{\mu\nu}+\Delta G_{\mu\nu}=8\pi GT_{\mu\nu},\label{nonlocal_field_eq}
\end{equation}
which in a Friedman-Lema\^{i}tre-Robertson-Walker (FLRW) background
\begin{equation}
ds^{2}=-dt^{2}+a^{2}d\vec{x}^{2},
\end{equation}
can be written as 
\begin{eqnarray}
3H^{2}+\Delta G_{00} & = & 8\pi G\rho,\\
-2\dot{H}-3H^{2}+\frac{1}{3a^{2}}\delta^{ij}\Delta G_{ij} & = & 8\pi Gp.\nonumber 
\end{eqnarray}
Here, the tensor $\Delta G_{\mu\nu}$ corresponds to the nonlocal
contribution and is given, for the FLRW metric, by the following expressions
\cite{deser_nonlocal_2007}, 
\begin{eqnarray}
\Delta G_{00}=\Bigl[3H^{2}+3H\partial_{t}]\Biggl\{ f\Bigr(\Box^{-1}R\Bigl)+\frac{1}{\square}\Bigl[Rf,\Bigr(\Box^{-1}R\Bigl)\Bigr]\Biggr\}+\frac{1}{2}\partial_{t}\Bigr(\Box^{-1}R\Bigl)\partial_{t}\bigg(\frac{1}{\square}\Bigl[Rf,\Bigr(\Box^{-1}R\Bigl)\Bigr]\bigg),\label{deltG00}\\
\Delta G_{ij}=a^{2}\delta_{ij}\Bigg[\frac{1}{2}\partial_{t}\Bigr(\Box^{-1}R\Bigl)\partial_{t}\bigg(\frac{1}{\square}\Bigl[Rf,\Bigr(\Box^{-1}R\Bigl)\Bigr]\bigg)-\Big(2\dot{H}+3H^{2}+2H\partial_{t}+\partial_{t}^{2}\Big)\left(f+\frac{1}{\square}\Bigl[Rf,\Bigr(\Box^{-1}R\Bigl)\Bigr]\right)\Bigg],\label{deltaGij}
\end{eqnarray}
where $\rho$ and $p$ are respectively the energy density and pressure
of a perfect fluid. From now on, a comma next to $f$ represents  a derivative of the function w.r.t its argument. Equations~(\ref{deltG00}-\ref{deltaGij}) can
be localized by introducing the auxiliary variables $X$ and $U$
defined as 
\begin{eqnarray}
\Box X &\equiv & R,\label{eq:boxx}\\
\Box U &\equiv & f_{,}R\label{eq:boxu}
\end{eqnarray}
With the use of the auxiliary functions $X$ and $U$, Eqs.~(\ref{deltG00}-\ref{deltaGij})
can be rewritten as 
\begin{eqnarray}
\Delta G_{00} & = & (3H^{2}+3H\partial_{t})(f+U)+\frac{1}{2}\dot{X}\dot{U},\\
\nonumber \\
\Delta G_{ij} & = & a^{2}\delta_{ij}\Bigg[\frac{1}{2}\dot{X}\dot{U}-\Big(2\dot{H}+3H^{2}+2H\partial_{t}+\partial_{t}^{2}\Big)\left(f+U\right)\Bigg].
\end{eqnarray}

The DW model has been shown to be capable of reproducing the background
evolution given by $\Lambda$CDM with $\Omega_{M}\approx0.28$ \cite{Deffayet:2009ca}
by fixing the nonlocal function to 
\begin{equation}
f(X)=0.245\Bigl[\tanh(0.350Y+0.032Y^{2}+0.003Y^{3})-1\Bigr],\label{fofY}
\end{equation}
with $Y\equiv X+16.5$. This choice fully determines the model and
no more free parameters are left.

The dependence of the nonlocal modification on $X$ is suggested by
quantum radiative corrections \cite{maggiore_perturbative_2016}
and is triggered mainly at the end of the radiation domination era
(since $R=0$ during radiation domination), with a slow evolution
afterwards. The interesting question arises then, whether the
DW model that gives the same background evolution as $\Lambda$CDM,
produces also the same behavior at perturbation level. The answer
is no, and in the following sections we will see why is it so. 

It is useful to write down modified Einstein equations (\ref{nonlocal_field_eq})
as well as auxiliary field equations (\ref{eq:boxx}-\ref{eq:boxu})
through $e$-folding time $N=\ln a$ 
\begin{eqnarray}
1+f+U+f'+U'+\frac{1}{6}X'U' & = & \Omega_{M}+\Omega_{R}\\
-2\xi-3+\frac{1}{2}X'U'-(2\xi+3)(f+U)-2(f'+U')-f''-U''-\xi(f'+U') & = & 3w_{M}\Omega_{M}+3w_{R}\Omega_{R}.
\end{eqnarray}
and
\begin{eqnarray}
X''+(3+\xi)X' & = & -RH^{-2}=-6\xi-12\label{eq:Xexpl}\\
U''+(3+\xi)U' & = & -Rf_{,}H^{-2}=-f_{,}(6\xi+12).\label{eq:Uexpl}
\end{eqnarray}
where $\xi\equiv H'/H$, $\Omega_{M}$ and $\Omega_{R}$ are the matter
and radiation fractional densities, respectively. We will use these
equations later.

\section{Perturbation equations}

\label{sec:perteq}

In this section, we introduce the linear scalar perturbation equations
for the DW model. Our method of getting perturbation equations is
similar to one implemented in Ref.~\cite{Koivisto:2008dh} and the
results are consistent up to some conventions. Here, we work in the
Newtonian gauge, in which scalar perturbations of the metric are given
by 
\begin{equation}
ds^{2}=-\left(1+2\Psi\right)dt^{2}+a^{2}\left(t\right)\left(1+2\Phi\right)\delta_{ij}dx^{i}dx^{j}
\end{equation}
We expand the auxiliary fields as $X+\delta X$ and $U+\delta U$.
In general for the anisotropic fluid in the first order of perturbation
we have 
\begin{eqnarray}
T_{0}^{0} & = & -\left(\rho+\delta\rho\right)\\
T_{i}^{0} & = & \left(\rho+p\right)v_{i}\\
T_{j}^{i} & = & \left(p+\delta p\right)\delta_{j}^{i}+\Sigma_{j}^{i}
\end{eqnarray}
Here we write the pressure perturbation $\delta p$ as $\delta p=c_{s}^{2}\delta\rho$,
where $c_{s}^{2}$ is the sound speed of the perfect fluid. The density
contrast $\delta$ is defined as $\delta\rho/\rho$ and $v_{i}$ is
the peculiar velocity field. In the case where the matter content
consists of radiation and non-relativistic matter, we have a vanishing
anisotropic stress tensor $\Sigma_{j}^{i}\simeq0$. Below we will
write down the linearly perturbed field equations

\begin{eqnarray}
\delta\left(G_{00}+\Delta G_{00}\right) & = & 8\pi G\delta T_{00},\\
\delta\left(G_{ij}+\Delta G_{ij}\right) & = & 8\pi G\delta T_{ij}\,,
\end{eqnarray}
 in Fourier space. The first order perturbation of the $(00)$ component
of Friedman equations is given by the following expression: 
\begin{eqnarray}
\delta G_{00} & = & 6H^{2}\Phi'+2\frac{k^{2}}{a^{2}}\Phi\\
\delta\Delta G_{00} & = & \frac{k^{2}}{a^{2}}f,\delta X+2\frac{k^{2}}{a^{2}}\Phi f+\frac{3}{2}H^{2}\left(X'\delta U'+U'\delta X'\right)\\
 & + & \frac{k^{2}}{a^{2}}\delta U+\frac{2k^{2}}{a^{2}}\Phi U\\
\delta T_{00} & = & \rho\delta
\end{eqnarray}
where the prime stands for derivative with respect to $e$-folding
time, $\ln a$. In the following equations we often put ourselves
in the sub-horizon limit ( $k/aH\gg1$). To do this, we assume that
$\Phi,\Psi,\delta U,\delta X,k^{-2}\delta$, and their $\ln a$ derivatives,
are all of the same order (as indeed can be verified a posteriori)
and systematically take the limit of large $k/aH$. For the $(ij)$
component, after contracting it with the projecting operator $(\dfrac{k^{i}}{k}\dfrac{k^{j}}{k}-\dfrac{1}{3}\delta^{ij})$
, we get 
\begin{equation}
\frac{2}{3}k^{2}\left(\Psi+\Phi\right)+\left(\dfrac{k^{i}}{k}\dfrac{k^{j}}{k}-\dfrac{1}{3}\delta^{ij}\right)\delta\Delta G_{ij}=-8\pi G\left(\rho+p\right)\sigma
\end{equation}
where $\sigma$ represents the anisotropic stress, and where $\delta\Delta G_{ij}$
in the sub-horizon limit is 
\begin{eqnarray}
\delta\Delta G_{ij} & = & D_{ij}\left(f,\delta X+\delta U\right)+\delta D_{ij}\left(f+U\right)\\
 & + & \frac{1}{2}H^{2}a^{2}\left(X'\delta U'+U'\delta X'\right)\delta_{ij}+H^{2}a^{2}\left(\Phi-\Psi\right)X'U'\delta_{ij}\nonumber 
\end{eqnarray}
where $D_{ij}$ and $\delta D_{ij}$ , also in the sub-horizon limit,
are respectively 
\begin{eqnarray}
D_{ij} & = & -\delta_{ij}k^{2}+k_{i}k_{j}\\
\delta D_{ij} & = & \left(-\delta_{ij}k^{2}+k_{i}k_{j}\right)\left(\Psi+\Phi\right).
\end{eqnarray}

Now to complete the set of equations we need also to perturb Eqs.~(\ref{eq:boxx}-\ref{eq:boxu}).
We get 
\begin{eqnarray}
 &  & \delta X''+\left(3+\xi\right)\delta X'+\hat{k}^{2}(\delta X+2\Psi+4\Phi)-2\Psi(X''+3X'+\xi X'+6+6\xi)\label{eq:perX}\\
 &  & -\Psi'(X'+6)+3\Phi'X'+6\Phi''+6\left(4+\xi\right)\Phi=0\nonumber 
\end{eqnarray}
and

\begin{eqnarray}
 &  & \delta U''+\left(3+\xi\right)\delta U'+\hat{k}^{2}\delta U-2\Psi U''-\left(2\left(3+\xi\right)\Psi+\Psi'-3\Phi'\right)U'\label{eq:perU}\\
 & = & -6f,,\delta X\left(\xi+2\right)+6f,\left(\Psi'+2\left(2+\xi\right)\Psi\right)\nonumber \\
 &  & -6f,\left(\Phi''+\left(4+\xi\right)\Phi\right)-2f,\hat{k}^{2}\left(\Psi+2\Phi\right).\nonumber 
\end{eqnarray}
where $\hat{k}=k/aH$. Moving to the sub-horizon limit we find, for
the $(00)$ component, 
\begin{eqnarray}
 &  & \Phi+\frac{f,\delta X}{2}+\Phi f+\frac{\delta U}{2}+\Phi U=4\pi G\frac{a^{2}\rho\delta}{k^{2}}\nonumber \\
 & = & \frac{3H_{0}^{2}}{2k^{2}}\left(\Omega_{R}^{0}a^{-2}\delta_{R}+\Omega_{M}^{0}a^{-1}\delta_{M}\right).\label{eq:subhor00}
\end{eqnarray}
For the $(ij)$ component, after acting with the projection operator,
we have 
\begin{equation}
\frac{2}{3a^{2}}k^{2}\left(\Psi+\Phi\right)+\frac{2}{3a^{2}}k^{2}[f,\delta X+\delta U+\left(\Psi+\Phi\right)\left(f+U\right)]=-8\pi G\left(\rho+p\right)\sigma,
\end{equation}
At late times, when the relativistic contribution is small, we can
neglect the contribution coming from the anisotropic stress, $\sigma\approx0$.
So we get 
\begin{equation}
\Psi+\Phi+f,\delta X+\delta U+\left(\Psi+\Phi\right)\left(f+U\right)=0\label{eq:subhorij}
\end{equation}
Eqs.~(\ref{eq:perX}-\ref{eq:perU}) reduce to 
\begin{eqnarray}
\delta X & = & -2\left(\Psi+2\Phi\right),\label{eq:subhorX}\\
\delta U & = & -2f,\left(\Psi+2\Phi\right).\label{eq:subhorU}
\end{eqnarray}
From the covariant conservation law of the energy-momentum tensor,
$\nabla^{\mu}T_{\mu\nu}=0$, we get finally the following equations
for the matter density perturbation $\delta_{M}$ in the sub-horizon
limit: 
\begin{eqnarray}
\delta_{M}''+\left(2+\xi\right)\delta_{M}' & = & -\hat{k}^{2}\Psi.\label{eq:growtheq}
\end{eqnarray}
In order to solve this equation we need to find an expression for
$\Psi$. This can be done by combining Eqs.~(\ref{eq:subhor00}-\ref{eq:subhorij})
and Eqs.~(\ref{eq:subhorX}-\ref{eq:subhorU}). After simple algebraic
manipulations we find for the modified gravity function $\eta$ and
for the potentials $\Psi$ and $\Phi$ the following expressions:
\begin{eqnarray}
\eta & = & \frac{\Phi+\Psi}{\Phi}=\frac{4f,}{1+U+f-4f,},\label{eq:eta}\\
\Psi & = & -\frac{3H_{0}^{2}\left(1+U-8f,+f\right)\Omega_{M}^{0}\delta_{M}}{2ak^{2}\left(1+U-6f,+f\right)\left(1+f+U\right)},\label{eq:psi}\\
\Phi & = & \frac{3H_{0}^{2}\left(1+U-4f,+f\right)\Omega_{M}^{0}\delta_{M}}{2ak^{2}\left(1+U-6f,+f\right)\left(1+f+U\right)}.\label{eq:phi}
\end{eqnarray}
Finally, by plugging the expression for $\Psi$ from Eq.~(\ref{eq:psi})
into Eq.~(\ref{eq:growtheq}), we obtain the $k$-independent growth
equation 
\begin{eqnarray}
\delta_{M}''+\left(2+\xi\right)\delta_{M}' & = & \frac{3H_{0}^{2}\left(1+U-8f,+f\right)\Omega_{M}^{0}\delta_{M}}{2a^{3}H^{2}\left(1+U-6f,+f\right)\left(1+f+U\right)}.\label{eq:growtheqfinal}
\end{eqnarray}

In order to solve numerically Eqs.~(\ref{eq:boxx}-\ref{eq:boxu}),
we set the following initial conditions deep inside radiation-dominated
period $(N_{in}=ln a_{in}^{*}=-16)$: 
\begin{equation}
X(a_{in}^{*})=U(a_{in}^{*})=X'(a_{in}^{*})=U'(a_{in}^{*})=0.\label{eq:initialcond}
\end{equation}
In Ref.~\cite{Woodard:2014iga} it was argued that these initial
conditions force the homogenous solutions of the localized model to
vanish, rendering it equivalent to the nonlocal versions of the DW
model. For the growth equation (\ref{eq:growtheq}), the initial conditions
deep into the matter era are taken to be as in pure CDM 
\begin{equation}
\delta_{M}(a_{in})=a_{in},\hspace{15mm}\frac{\delta_{M}'(a_{in})}{\delta_{M}(a_{in})}=1,
\end{equation}
where the initial scale factor $a_{in}$ is taken at redshift $z_{in}=9$.
In the next section we generalize the initial conditions.

\begin{figure}[tbh]
\includegraphics[width=8.1cm]{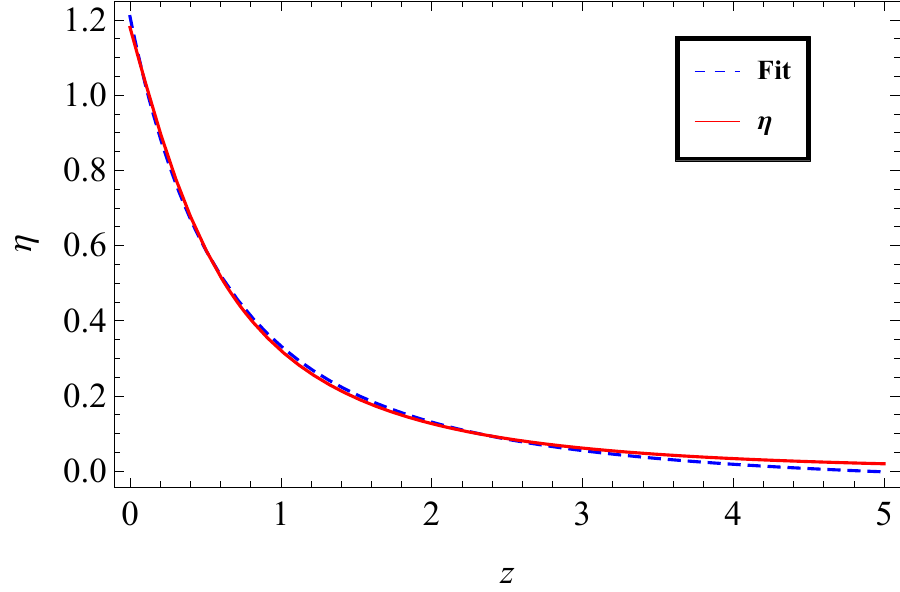} \includegraphics[width=8.2cm]{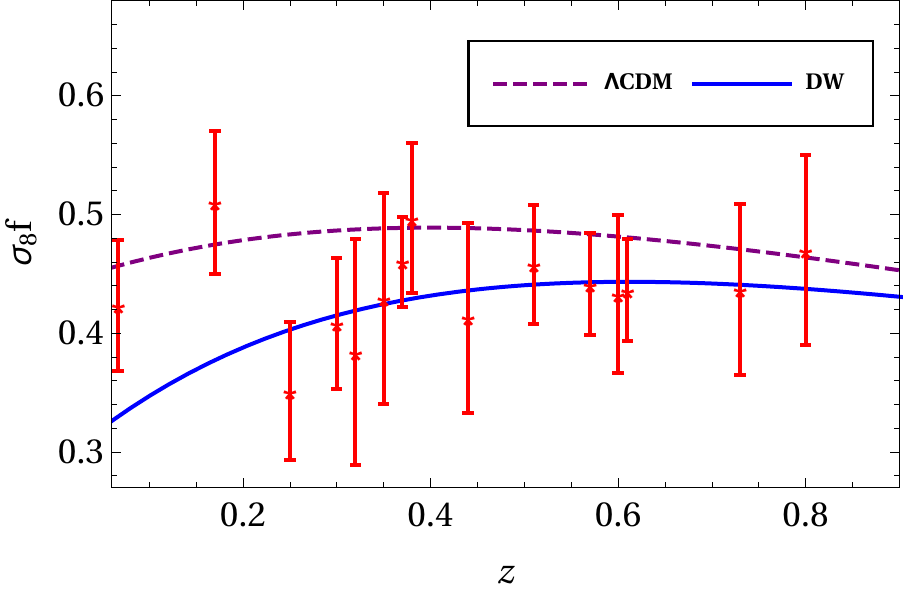}
\protect\caption{\textit{Left panel:} The evolution of anisotropic stress $\eta$ as
a function of redshift and best fit with the polynomial $A_{3}a^{3}+A_{2}a^{2}+A_{0}$,
where $\lbrace A_{3},A_{2},A_{0}\rbrace=\lbrace-0.54,1.79,-0.05\rbrace$.
\textit{Right panel:} The growth rate $f\sigma_{8}(z)$ for $\Lambda$CDM
and DW as functions of redshift.}

\label{ourplots} 
\end{figure}

The numerical results for the anisotropic stress $\eta$ and the growth
rate $f\sigma_{8}(z)\equiv\sigma_{8}\delta_{M}'/\delta_{M}$ fixing
$\Omega_{M}^0=0.3$ (which can be directly extracted from the observational
RSD data, see~\ref{tab:data}) are presented in Figure~\ref{ourplots}.
The  DW model fits the data better than $\Lambda$CDM ($\chi^2$ per d.o.f. equal to 0.736 instead of 0.943,
see Tab. \ref{tab:results}). The resulting lower normalization, $\sigma_8=0.78$, is in agreement 
to within 1$\sigma$ with the recent estimates based on
lensing \cite{Hildebrandt:2016iqg} ($\sigma_{8}=0.745_{-0.038}^{+0.038}$ for $\Omega_M^0=0.3$, Table $F_{2}$ - column $S_{8}$), contrary to $\Lambda$CDM. See also \cite{Ade:2015rim}.

Our results differ significatively from those in \cite{Dodelson:2013sma},
in which $f\sigma_{8}(z)$ lies above the $\Lambda$CDM curve, although
we agree with their results at the background level. We have not been
able to point out the reasons for this discrepancy. One could conjecture
it might be due to the fact we are solving a localized version of
the model in which the solution depend on the initial conditions on
$\delta X,\delta U$, which in the sub-horizon limit are not free
quantities but depend on $\Phi,\Psi$, and therefore on $\delta_{M}$.
However, the quasi-static solution is the attractor solution for large
$k$, so it is unlikely that the discrepancy depends on the localization
procedure or the quasi-static approximation. Our expression (\ref{eq:subhorX})
is indeed different from the corresponding quasi-static non-local expression Eq.
(29) in \cite{park_structure_2013}, but the two expressions coincide asymptotically in time in the limit
in which $\Psi,\Phi$ are weakly time-dependent with respect to the
rapidly varying sinusoidal function in the integrand for large
$k$, conditions that are met in the quasi-static limit. It is still possible that the disagreement is due
to the asymptotic equivalence between the local and non-local quasi-static limits not having been reached by the present time.

The behavior of $\eta$ can be approximated with an analytic fit of
the form: $A_{3}a^{3}+A_{2}a^{2}+A_{0}$, where $A_{3}$, $A_{2}$
and $A_{0}$ are free parameters. We find the following best fit results
with a percentage error up to $2\%$ for the quantity $1+\eta$ in
the range $z\in(0,5)$: $\lbrace A_{3},A_{2},A_{0}\rbrace=\lbrace-0.54,1.79,0.95\rbrace$
(see Figure~\ref{ourplots}, left panel). 

\begin{figure}[tbh]
\includegraphics[width=8.5cm]{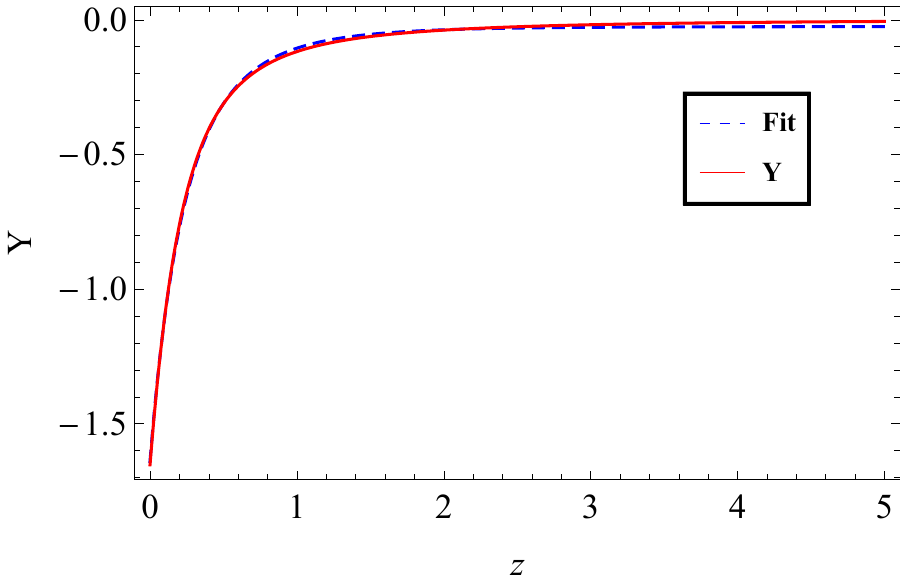} \includegraphics[width=8.5cm]{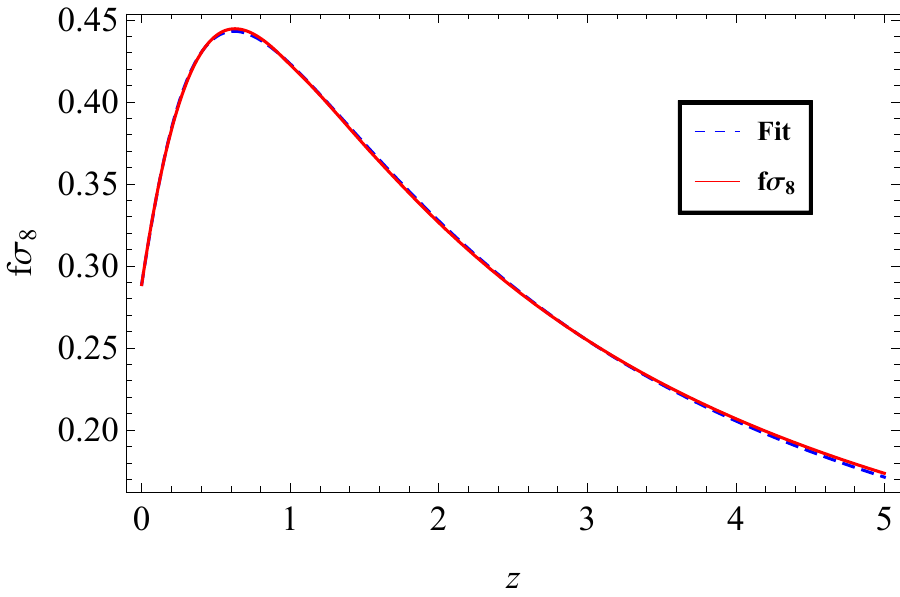}\protect\caption{\textit{Left panel:} The evolution of the effective gravitational
constant correction $Y=G_{eff}-1$ as a function of red-shift and
best fit with the polynomial $A_{4}a^{4}+A_{3}a^{3}+A_{0}$, where
$\lbrace A_{4},A_{3},A_{0}\rbrace=\lbrace-1.95,0.33,-0.03\rbrace$.\textit{
Right panel:} The evolution of $f\sigma_{8}(z)$ as a function of
redshift and best fit with the polynomial $A_{4}a^{4}+A_{3}a^{3}+A_{2}a^{2}+A_{1}a$,
where $\lbrace A_{4},A_{3},A_{2},A_{1}\rbrace=\lbrace1.41,-3.03,0.97,0.94\rbrace$.}

\label{Y} 
\end{figure}

We have also checked the ghost-freeness condition for the localized
theory derived in Ref.~\cite{Koivisto:2008dh} 
\begin{equation}
6f,>1+f+U>0,\label{ghostfree}
\end{equation}
and we found that at all times the condition $6f,>1+f+U$ is violated,
while the condition $1+f+U>0$ is always satisfied. In Ref.~\cite{Zhang:2011uv} the authors have discussed a particular case when the ghost-free condition~(\ref{ghostfree}) is satisfied and leads to an interesting cosmology. In the case of tensor-scalar theories it has been shown that the appearance of a
ghost mode in the theory's spectrum will lead to a situation where
the effective gravitational constant $G_{\mathrm{eff}}$ (in units
of $G_{\mathrm{Newton}}$) defined as 
\begin{equation}
G_{\mathrm{eff}}=1+Y=-\frac{2k^{2}\Psi}{3(aH)^{2}\Omega_{M}^{0}\delta_{M}}=\frac{(1+U-8f,+f)}{(1+U-6f,+f)(1+U+f)}
\end{equation}
will become negative \cite{amendola_phantom_2004}. From Figure~\ref{Y},
left panel, we indeed see that $Y$ is always negative when the non-local
contributions are non negligible. This explains why the perturbations
grow more slowly than in $\Lambda$CDM.

We see that $G_{\mathrm{eff}}$ goes negative near the present epoch.
This however is only true in our linear approximation. Near non-linear
structures, one must assume the existence of a screening mechanism
in order to pass local gravity constraints, so that within the screening
radius, standard gravity is recovered. This issue has been already
discussed in Ref.~\cite{Woodard:2014iga} and we refer to that paper
for further information. As it was done for the case of $\eta$, the
behaviors of $Y$ and $f\sigma_{8}(z)$ can be approximated with analytical
fits of the polynomial form $A_{4}a^{4}+A_{3}a^{3}+A_{0}$. The best
fit result in the case of $Y$, with a percentage error up to $3\%$
for the quantity $1+Y$ within $z\in(0,5)$, corresponds to $\lbrace A_{4},A_{3},A_{0}\rbrace=\lbrace-1.95,0.33,0.97\rbrace$
(see Figure~\ref{Y}, left panel). For the case of $f\sigma_{8}(z)$
we find the best fit result $A_{4}a^{4}+A_{3}a^{3}+A_{2}a^{2}+A_{1}a$.
with a percentage error up to $1\%$ in the same redshift range, for
$\lbrace A_{4},A_{3},A_{2},A_{1}\rbrace=\lbrace1.41,-3.03,0.97,0.94\rbrace$
(see Figure~\ref{Y}, right panel).

\section{Sound speed}

\label{appendix1} In order to ascertain that the quasi-static approximation
is valid, one has to ensure the absence of gradient instabilities,
i.e. whether the sign of the sound-speed of the perturbative-quantities
$\delta X$ and $\delta U$ is positive. To do this, we rewrite Eqs.~(\ref{eq:perX}-\ref{eq:perU})
in the following way:

\begin{eqnarray}
\delta X''+3\delta X'+\hat{k}^{2}(\delta X+2\Psi+4\Phi) & = & 0\label{eq:perX1}\\
\delta U''+3\delta U'+\hat{k}^{2}(\delta U+2f,\Psi+4f_{,}\Phi) & = & 0\label{eq:perU1}
\end{eqnarray}
keeping now the derivatives of $\delta X,\delta U$, but still neglecting
$\Psi,\Phi$ and their derivatives with respect to $\hat{k}^{2}\Psi,\hat{k}^{2}\Phi$.
Now from Eqs.~(\ref{eq:subhor00}-\ref{eq:subhorij}) in the case
of vacuum ($\delta=\sigma=0$) we find that 
\begin{equation}
\Phi=\Psi=-\frac{f,\delta X+\delta U}{2(1+f+U)}.\label{eq:pervacuum}
\end{equation}
So, by inserting Eq.~(\ref{eq:pervacuum}) into Eqs.~(\ref{eq:perX1}-\ref{eq:perU1})
we get finally 
\begin{equation}
\delta X''+3\delta X'+\hat{k}^{2}\delta X=\frac{3\hat{k}^{2}(f,\delta X+\delta U)}{(1+f+U)}
\end{equation}
and

\begin{equation}
\delta U''+3\delta U'+\hat{k}^{2}\delta U=\frac{3f,\hat{k}^{2}(f,\delta X+\delta U)}{(1+f+U)}
\end{equation}
We can combine these two equations and write them in a matrix form
\begin{equation}
X_{i}''+3X_{i}'+\hat{k}^{2}S_{ij}X_{j}=0,
\end{equation}
where $S_{ij}$ is a two by two matrix, defined as 
\begin{equation}
S=\frac{1}{(1+f+U)}\left(\begin{smallmatrix}1+f+U-3f, & -3\\
-3f,^{2} & 1+f+U-3f,
\end{smallmatrix}\right)
\end{equation}
and $X_{i}$ are the components of the vector $X=(\delta X,\delta U)$.
If $S$ is positive-definite, the perturbative quantities $\delta X$
and $\delta U$ have a positive sound-speed independent of the propagation
direction. An arbitrary matrix $A$ is called positive-definite when
it has only positive eigenvalues. The eigenvalues of the matrix $S$ are 
\begin{equation}
\lambda_{1} = 1, \hspace{15mm} \lambda_{2}  = \frac{\left(1+f+U-6f,\right)}{\left(1+f+U\right)}.
\end{equation}\label{eq:eigenvalues}
As mentioned in the previous section, for the DW model one has $(1+f+U)>0$. On the other hand from the violation of the ghost-free condition (\ref{ghostfree}) we have that $(6f,-f-U-1)<0$. Under these two conditions both eigenvalues of the matrix $S$ are always positive and so the matrix $S$ is positive-definite. Here, we conclude that the quasi-static
approximation is a valid one, which means that the solution based
on this approximation is an attractor one and any solution of Eqs.~(\ref{eq:perX}-\ref{eq:perU})
should approach it at some point. In Ref. \cite{Koivisto:2008dh},
the same procedure is carried out in the Einstein frame, with the
same result.

\begin{table}[t]
\centering{} %
\begin{tabular}{|c|c|c|c|}
\hline 
Survey  & $z$  & $\sigma_{8}f$  & References\tabularnewline
\hline 
\hline 
6dFGRS  & $0.067$  & $0.423\pm0.055$  & Beutler et al. (2012)~\cite{2012MNRAS.423.3430B}\tabularnewline
\hline 
\multirow{2}{*}{LRG-200} & $0.25$  & $0.3512\pm0.0583$  & \multirow{2}{*}{Samushia et al. (2012)~\cite{Samushia:2011cs}}\tabularnewline
\cline{2-3} 
 & $0.37$  & $0.4602\pm0.0378$  & \tabularnewline
\hline 
\multirow{5}{*}{BOSS} & $0.30$  & $0.408\pm0.0552$  & \multirow{2}{*}{Tojeiro et al. (2012)\cite{boss_tojeiro_2012}}\tabularnewline
\cline{2-3} 
 & $0.60$  & $0.433\pm0.0662$  & \tabularnewline
\cline{2-4} 
 & $0.38$  & $0.497\pm0.063$  & \multirow{3}{*}{Alam et al. (2016) \cite{Alam:2016hwk}}\tabularnewline
\cline{2-3} 
 & $0.51$  & $0.458\pm0.050$  & \tabularnewline
\cline{2-3} 
 & $0.61$  & $0.436\pm0.043$  & \tabularnewline
\hline 
\multirow{2}{*}{WiggleZ } & $0.44$  & $0.413\pm0.080$  & \multirow{2}{*}{Blake (2011) \cite{Blake:2012pj}}\tabularnewline
\cline{2-3} 
 & $0.73$  & $0.437\pm0.072$  & \tabularnewline
\hline 
Vipers  & $0.8$  & $0.47\pm0.08$  & De la Torre et al. (2013) \cite{vipers_delatorre_2013}\tabularnewline
\hline 
2dFGRS  & $0.17$  & $0.51\pm0.06$  & Percival et al. (2004) \cite{Percival:2004fs,Song:2008qt}\tabularnewline
\hline 
LRG  & $0.35$  & $0.429\pm0.089$  & Chuang and Wang (2013) \cite{chuang_wang_2013}\tabularnewline
\hline 
LOWZ  & $0.32$  & $0.384\pm0.095$  & Chuang et al. (2013) \cite{Chuang:2013wga}\tabularnewline
\hline 
CMASS  & $0.57$  & $0.441\pm0.043$  & Samushia et al. (2013) \cite{Samushia:2013yga}\tabularnewline
\hline 
\end{tabular}\protect\protect\caption{\label{tab:data} Up-to-date RSD measurements from various sources.
These are the points shown in Fig. (\ref{ourplots}) and Fig. (\ref{2dlike_growth}).}
\end{table}



\section{Model-independent constraints}

Eq.~(\ref{eq:growtheq}) is a second order differential equation
for the density contrast $\delta_{M}$ and, in order to solve it,
we need to specify two initial conditions. The typical choice corresponds
to a standard cosmology dominated by pressureless matter at high redshifts,
in which $\delta_{M}\sim a$. Namely, one assumes 
\begin{equation}
\delta_{M}(a_{in})=Aa_{in},\hspace{15mm}\frac{\delta_{M}'(a_{in})}{\delta_{M}(a_{in})}=1,
\end{equation}
where $a_{in}$ is some arbitrary initial value of the scale factor
$a$ outside the range of redshift for which we have observations,
say at redshift $z=9$.

These initial conditions, however, depend on several assumptions about
the past: they require, in fact, that matter dominates ($\Omega_{M}=1$),
that matter is pressureless, that any decaying mode has been suppressed,
and that gravity is Einsteinian. Broadly speaking, there is very little
direct proof for any of these assumptions. Let us consider for instance
two analytical toy models. In the first, one can imagine that there
is a fraction $\Omega_{h}$ of a homogeneously distributed component
along with matter in the past, just like in models of Early Dark Energy
(except we are not requiring this component to lead to acceleration
at the present). Then the growth of fluctuation obeys the equation
\begin{equation}
\delta_{M}''+\left(2+\xi\right)\delta_{M}'-\frac{3}{2}\Omega_{M}\delta_{M}=0,\label{eq:fluc-eq}
\end{equation}
with $\Omega_{M}=1-\Omega_{h}$ instead of $\Omega_{M}=1$. In this
case, the growth exponent is no longer $\delta_{M}\sim a^{1}$ but
rather $\sim a^{p}$ where 
\begin{equation}
p=\frac{1}{4}(-1\pm\sqrt{1+24\Omega_{M}})
\end{equation}
Therefore, if for instance $\Omega_{h}=0.05$, a value consistent
with the analysis in \cite{2013PhRvD..87h3009P} the total growth from
$z_{CMB}\approx1100$ to $z_{0}=0$ is smaller than the corresponding
pure CDM one (we are neglecting here the final accelerated epoch)
by a factor 
\begin{equation}
\frac{(z_{CMB})^{p}}{z_{CMB}}\approx0.8
\end{equation}
So the existence of a small non-vanishing homogeneous component would
produce a value of $\sigma_{8}$ which would be $0.8$ times smaller
than the Planck $\Lambda$CDM value.

The second analytical toy case comes from the simplest Brans-Dicke
model, parametrized by the Brans-Dicke coupling parameter $\omega$.
In such a case in fact one has during the matter era 
\begin{equation}
\frac{\delta_{M}'(a_{in})}{\delta_{M}(a_{in})}=\frac{2+\omega}{1+\omega}
\end{equation}
rather than unity. If the Brans-Dicke gravity is universal and unscreened,
then $\omega\gg1$ because of local gravity constraints, and one recovers
the standard initial condition. But if the scalar force is not universal
and baryons are uncoupled or, alternatively, if the force is screened
by a chameleon-like mechanism, then $\delta_{M}'/\delta_{M}$ can
deviate substantially from standard.

These two toy models show that if one wants to test modified gravity,
and not also at the same time the entire CDM paradigm, then one
needs to isolate the effects of modified gravity from those that depend
on different assumptions. The simplest way to do so is to introduce
then two new parameters that correspond to the two initial values
of the growth equation (\ref{eq:fluc-eq}) and marginalize the likelihood
over them. Instead of $\delta_{M}(a_{in})$ we adopt the present normalization
$\sigma_{8}^{0}$ as first parameter, and 
\begin{equation}
\alpha\equiv\frac{\delta_{M}'(a_{in})}{\delta_{M}(a_{in})}
\end{equation}
as second free parameter. It is worth mentioning that our approach
is of course not completely model independent, in the sense that we
still assume that the matter content in the observational range is
given by a pressureless perfect fluid and is conserved.

The quantity of interest in this paper is the RSD observable, also
called growth rate: 
\begin{eqnarray}
 & f\sigma_{8}(z)\equiv\sigma_{8}(\ln\delta_{M})',
\end{eqnarray}
with the amplitude of fluctuations $\sigma_{8}(z)$ defined as 
\begin{eqnarray}
 & \sigma_{8}(z)=\sigma_{8}^{0}\frac{\delta_{M}(z,k)}{\delta_{M}(0,k)}.
\end{eqnarray}
The current value of the amplitude of fluctuations $\sigma_{8}^{0}$
is estimated through e.g. weak lensing \cite{More:2014uva}, the
cosmic microwave background power spectrum \cite{Ade:2015xua} or
cluster abundances \cite{Ade:2015fva}. In all these cases the estimate
depends, in general, on the choice of a gravity theory. As a consequence
the current value of $\sigma_{8}^{0}$ is highly model-dependent.

\section{Likelihood analysis}

\label{sec:results}

Assuming that $\sigma_{8}^{0}$ and $\alpha$ can take any value for
modified gravity models, it is of interest to see how the growth rate
$f\sigma_{8}(z)$ behaves, in terms of agreement with data, when we
allow those two parameters to vary. In what follows we always fix
$\Omega_{M}=0.3$ for simplicity.

In general, the likelihood function of a given model (represented
by the parameter vector $\Theta$) with respect to some data is given
by $L(\Theta)=A\exp[-\chi^{2}/2]$ 
with
\begin{eqnarray}
\chi^{2}=(D-T)^{T}C^{-1}(D-T).\label{eq:chi2}
\end{eqnarray}
Here, $D$ and $T$ are respectively the data and theory vectors,
$C$ is the covariance matrix and $A$ is a normalization constant.
The vector $D$ contains the measurements of the observable quantity
(in our case, the growth rate $f\sigma_{8}(z)$) for each point (i.e.
each redshift value) and $T$ represents the corresponding predictions
for that observable. When the data points are independent, $\chi^{2}$
takes the form 
\begin{eqnarray}
\chi^{2}=\sum_{i}\frac{(d_{i}-t_{i})^{2}}{\sigma_{i}^{2}},\label{eq:chi2indep}
\end{eqnarray}
so the likelihood function may be replaced by the simpler expression 
\begin{eqnarray}
L(\Theta)=A\exp[-\frac{1}{2}\sum_{i}\frac{(d_{i}-t_{i})^{2}}{\sigma_{i}^{2}}],\label{eq:indeplike}
\end{eqnarray}
In (\ref{eq:indeplike}), $d_{i}$ and $t_{i}$ are respectively the
elements of $D$ and $T$, and $\sigma_{i}$ is the error associated
to the measurement $d_{i}$.

We assume our data points (taken from \cite{Alam:2016hwk}) to be
independent and thus use (\ref{eq:chi2indep}, \ref{eq:indeplike}). We choose to integrate
the equation for density perturbations (\ref{eq:growtheq}) from the
redshift $z_{in}=9$. The results of the fits are shown in Table \ref{tab:results},
in which we display the $\chi^{2}$ deviation given by $\Lambda$CDM,
by the DW model and by the DW varying first only $\sigma_{8}^{0}$
(Case I) and then varying both $\sigma_{8}^{0},\alpha$ (Case II). 

\begin{table}[t]
\centering{} %
\begin{tabular}{|c|c|c|c|}
\hline 
Model  & $\sigma_{8}^{0}$  & $\alpha$  & $\chi^{2}/\mathrm{dof}$ \tabularnewline
\hline 
\hline 
$\Lambda$CDM  & $0.83$  & $1$  & $0.943$\tabularnewline
\hline 
DW  & $0.78$  & $1$  & $0.736$\tabularnewline
\hline 
DW: case I  & $0.81\pm0.03$  & $1$  & $0.696$\tabularnewline
\hline 
DW: case II  & $1.06\pm0.28$  & $-1.27\pm0.1$  & 0.668\tabularnewline
\hline 
\end{tabular}\protect\protect\caption{\label{tab:results} The $\chi^{2}/\mathrm{dof}$ for RSDs measurements
for $\Lambda$CDM, the DW and for the best-fit cases of the DW model
for $\sigma_{8}^{0}$ (Case I) and $\{\sigma_{8}^{0},\alpha\}$ (Case
II), respectively.}
\end{table}

\begin{figure}[tbh]
\includegraphics[width=8.5cm]{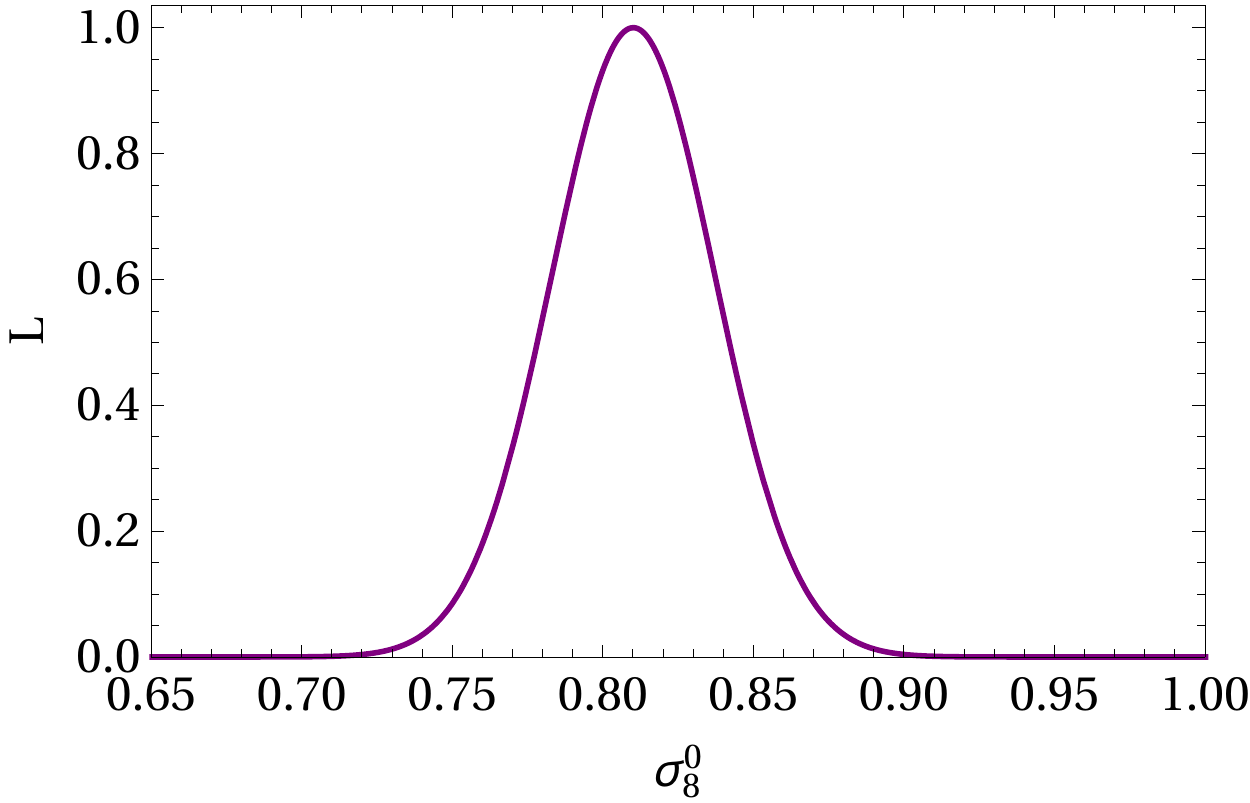} \protect\caption{The likelihood for $\sigma_{8}^{0}$, when $\text{\ensuremath{\alpha}}$
is fixed to $1$ (Case I). The maximum lies at $\sigma_{8}^{0}=0.81$.}

\label{1dlike} 
\end{figure}

The Case I likelihood for $\sigma_{8}^{0}$ is displayed in Figure~(\ref{1dlike}).
For both Case I and Case II, the best-fit yields a better performance
compared with $\Lambda$CDM, but the $\chi^{2}$ improves only marginally
upon the no-free parameters DW.

\begin{figure}[tbh]
\includegraphics[width=8.5cm]{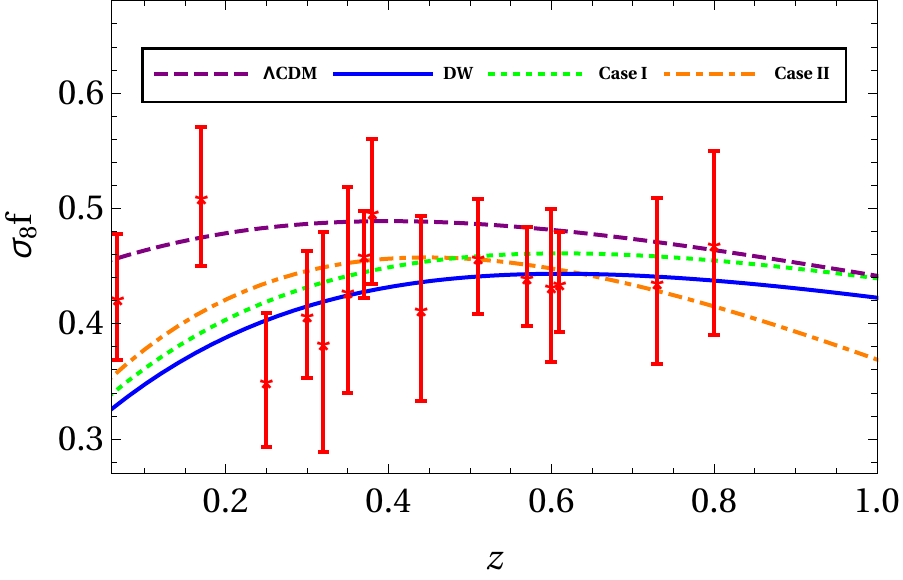} \protect\caption{ The growth rate $f\sigma_{8}(z)$ for $\Lambda$CDM and the three
best-fit cases for the DW model. The data points are collected in
Table \ref{tab:data}.}

\label{2dlike_growth} 
\end{figure}

In Figure~(\ref{2dlike_growth}) we show the growth rate given by
the three analysis and by $\Lambda$CDM.

\section{Conclusion}

\label{sec:conclusion}

In this work, we have extended the analysis performed in \cite{Dodelson:2013sma}
regarding the predictions of RSD given by the DW model of nonlocal
gravity. We have found that the localized version of the theory is
not ruled out by the RSD data, as was the case for the previous analysis,
and actually leads to a better agreement than the standard $\Lambda$CDM
cosmology. Ultimately, this behavior is due to the violation of the
ghost-free condition. In Ref.~\cite{Maggiore:2016gpx} it has been
argued that the existence of ghosts in non-local theories does not
necessarily rule out the model since the ghost mode is actually ``frozen''
due to fixed boundary conditions imposed on the model. Interestingly,
the predicted value $\sigma_{8}=0.78$ for $\Omega_{M}^{0}=0.3$ is in
agreement to within 1$\sigma$ with the recent estimates ($\sigma_{8}=0.745_{-0.038}^{+0.038}$)  based on
lensing, see e.g. \cite{Ade:2015rim,Hildebrandt:2016iqg}. We have
also investigated how much the fit improves when we generalize the
initial conditions of the growth rate function, and we find that the
improvement is just marginal.

Our perturbation results do not agree with the analysis in \cite{Dodelson:2013sma},
who integrated numerically the equations in their nonlocal form.
This could be due to an intrinsic difference between the nonlocal
and the localized versions of the DW model, for instance in the way
the quasi-static limit is performed. Unfortunately, we have been unable
to point out the reason for this discrepancy, notwithstanding extended
testing. In any case, we believe the localized version of the DW theory
gives interesting predictions on linear perturbation level and deserves
further consideration. \acknowledgments We thank S. Doldelson, S.
Park, Y. Akrami, J. Rubio, T. Koivisto, V. Pettorino, and S. Casas
for several useful discussions. We acknowledge support from DFG through
the project TRR33 ``The Dark Universe.'' H.N. acknowledges financial
support from DAAD through the program ``Forschungsstipendium f{\"u}r
Doktoranden und Nachwuchswissenschaftler.''

 \bibliographystyle{apsrev}
\bibliography{amendolamodnonloc,Henrik,nonlocalRefsTSK}

\end{document}